\documentclass[twocolumn,floatfix]{revtex4-2}
\pdfoutput=1
\usepackage{graphicx}
\usepackage{dcolumn}
\usepackage{longtable}
\usepackage{amsmath}
\usepackage{amssymb}
\usepackage{color}
\usepackage{isotope}

\begin{document}
\title{Islands of shape coexistence for ${\bf Z=38}$-84 in a non-relativistic mean-field approach using Hartree-Fock-Bogoliubov theory}

\author
{Malik A. Hasan$^1$ and Dennis Bonatsos$^2$}

\affiliation
{$^1$  Department of Physics, College of Science, University of Kirkuk, Kirkuk, Iraq}

\affiliation
{$^2$Institute of Nuclear and Particle Physics, National Centre for Scientific Research ``Demokritos'', GR-15310 Aghia Paraskevi, Attiki, Greece}

\begin{abstract}

Based on the microscopic mechanism of the particle-hole (p-h) excitations in the proton and neutron single-particle energy levels relative to the Fermi energy, a search for islands of shape coexistence (SC) is performed over a wide range of even-even nuclei from $Z=38$ to 84 using non-relativistic self-consistent mean-field with the Hartree-Fock-Bogoliubov (HFB) theory using the Skyrme-SKI3 functional. The results of the present study show that neutron-induced islands of SC, corresponding to proton p-h excitations, are found around the magic numbers $Z=82$ and $Z=50$, centered at the relevant neutron midshells of $N=104$ and $N=66$ respectively, while proton-induced islands of SC, corresponding to neutron p-h excitations, are found around the neutron numbers $N=90$ and $N=60$, centered at the relevant proton midshells $Z=66$ and $Z=38$ respectively. In addition, islands of SC due to both neutron and proton particle-hole excitations are found around $N=40$, $Z=40$. The results of the present study are compared with the results of covariant density functional theory using the DDME2 functional, using the   same p-h mechanism. The islands of SC that appeared in the CDFT work with the DDME2 functional are corroborated by the present study with the Skyrme-SKI3 functional, thus confirming the robustness of the particle-hole excitations mechanism in searching for islands of SC. In addition, the current study revealed new regions of SC, adjacent to the earlier islands and expanding their shores. 
 
\end{abstract}

\maketitle

\section{Introduction}

The nature of nuclear interactions between the constituents of the nucleus leads to the emergence of interesting phenomena that attract researchers to explore them through theoretical and experimental methods. One of the unique phenomena in a finite many-body quantum system, appearing in nuclei near closed shells, is the coexistence of different equilibrium shapes at similar energies, known as shape coexistence (SC) (see \cite{Heyde1983,Wood1992,Heyde2011,Garrett2022,Bonatsos2023c,Leoni2024} for relevant reviews). Thanks to the rapid development worldwide of rare isotope radioactive ion beam facilities \cite{Yano2007,Glasmacher2017}, SC has become one of the important topics that have attracted the attention of researchers in recent years, looking in particular for experimental evidence for new regions of SC,  as well as for confirmation of the existence of these regions through  relevant theoretical studies. 
 
SC has been first suggested by Morinaga  in 1956 in the spectrum of oxygen nuclei ($^{16}$O) \cite{Morinaga1956}. SC  is said to occur in a nucleus if the ground state band is accompanied by another similar band possessing the same angular momenta at similar energies, but with the two bands having radically different structure, for example with one of them being spherical and the other one deformed. Over the years, studies have shown that SC is an effect that can appear in many regions of the nuclear chart \cite{Heyde2011,Garrett2022,Bonatsos2023c,Leoni2024}. Extended regions of SC around the magic numbers $Z=82$ and $Z=50$, as well as around $N=90$, 60, 40 and $Z=64$, 40, 34 have been seen, as depicted schematically  in Fig. 8 of Ref. \cite{Heyde2011}.

On the theoretical front, particle-hole excitations across magic shell closures have been considered since the early days of SC as the main microscopic mechanism behind SC \cite{Heyde1985,Heyde1989,Heyde1990a,Heyde1990b}, with configuration mixing \cite{Duval1981,Duval1982} of the ground state band (which has no p-h excitations) and the excited configuration (which possesses p-h excitations) leading to two coexisting bands \cite{Fortune2017,Fortune2019a,Fortune2019b}. The concept of p-h excitations led to the development of the Interacting Boson Model with configuration mixing (IBM-CM) \cite{DeCoster1996,Lehmann1997,DeCoster1997,DeCoster1999,GarciaRamos2011,GarciaRamos2014a,GarciaRamos2014b,GarciaRamos2015}, within which a unified description of SC and shape phase transitions (SPT) has been obtained  \cite{Heyde2004,GarciaRamos2018,GarciaRamos2019,GarciaRamos2020,MayaBarbecho2022}.
A unified description of SC and SPT has also been achieved in the framework of the phenomenological Bohr-Mottelson model 
\cite{Budaca2017,Budaca2019,Buganu2019,AitBenMennana2021,AitBenMennana2022,Benjedi2024,Buganu2025}. 

More recently, extensive mean field calculations both in the non-relativistic \cite{Bender2003} and relativistic \cite{Reinhard1989,Ring1996,Vretenar2005} frameworks have become possible, and relevant codes have been developed (see Refs. \cite{Maruhn2014,Reinhard2021} and \cite{Ring1997,Niksic2014} respectively). Extensive mean field calculations have been performed along several series of isotopes, in order to study the evolution of the nuclear shapes and the appearance of SC (see Refs. \cite{Niksic2002,Meng2005,Nomura2011a,Nomura2011b,Nomura2012,Nomura2013,Nomura2016,Nomura2017,Kumar2021,Nomura2022}, for example).

Furthermore, SC phenomena in all mass regions have been investigated using a dual-shell mechanism \cite{Martinou2021,Martinou2023,Bonatsos2023b} based on the Elliott SU(3) \cite{Elliott1958a,Elliott1958b,Elliott1963,Harvey1968} and the proxy-SU(3) symmetry \cite{Bonatsos2023a}. This mechanism is based on the interplay between the Harmonic-Oscillator (HO) magic numbers 2, 8, 20, 40, 70, 112, 168, and the spin-orbit (SO) like magic numbers 6, 14, 28, 50, 82, 126, 184 through the particle excitations occurring between the HO and SO sets of shells. The main prediction of the dual shell mechanism is that SC of the ground state band and a nearby $0^+$ excited band cannot occur everywhere, but only within the nucleon (proton or neutron) numbers 7-8, 17-20, 34-40, 59-70, 96-112, 146-168 \cite{Martinou2021}, in rough agreement with the empirical regions shown in Fig. 8 of the review article \cite{Heyde2011}, for which the term \textit{islands of shape coexistence} has been coined. 

The predictions of the dual shell mechanism have been tested against microscopic calculations within the covariant density functional theory approach, using the DDME2 functional \cite{Bonatsos2022a,Bonatsos2022b}, which confirmed the presence of islands of SC, further clarifying the role played by the proton-neutron interaction in their formation.
In the case of the proton-induced SC, the neutron p-h excitations are caused by the protons, which act as elevators of the neutrons through the proton-neutron interaction, creating holes in the proton orbitals, while in the case of the neutron-induced SC, the proton p-h excitations are caused by the neutrons, which act as elevators of the protons through the neutron-proton interaction, creating holes in the neutron orbitals \cite{Bonatsos2022b}. Neutron-induced SC is seen, for example, around the magic numbers $Z=50$ and 82, in agreement with the early explanation of SC in these regions as due to p-h excitations \cite{Heyde1985,Heyde1989,Heyde1990a,Heyde1990b}, while proton-induced SC is seen, for example, in the regions $N=90$, $Z=64$ and $N=60$, $Z=40$ \cite{Bonatsos2022a}. The question on the extent in which the predictions of the dual shell mechanism are supported by non-relativistic mean-field calculations remained open \cite{Bonatsos2022b}. 

The main objective of the present work is to investigate to what extent islands of SC occur for $Z=38$-84,  based on the particle-hole (p-h) excitation mechanism in the single-particle energy level ($E_{sp}$) of proton and neutron relative to the Fermi energy ($E_F$) using a non-relativistic self-consistent mean-field code HFBTHO \cite{Stoitsov2013} with the Hartree-Fock-Bogoliubov (HFB) theory using Skyrme-SKI3 functional \cite{Reinhard1995}. More specifically, we examine how the relative position of orbitals that belong to certain spherical shells changes with respect to the Fermi surface for a series of isotopes or isotones. States located below the Fermi surface correspond to particles, while states appearing above it correspond to holes. Following a series of isotopes, we observe if there are proton states that change from particles to holes and vice versa. When this happens for a certain range of neutrons we propose that in that section neutron-induced shape coexistence occurs. By analogy, following a series of isotones, we observe if there are neutron states that change from particles to holes and vice versa. When this happens for a certain range of protons we propose that in that section proton-induced shape coexistence occurs. The results of the present study, which is using non-relativistic mean-field HFB theory, will then be compared to the results of the covariant density functional theory with the DDME2 functional of Ref.\cite{Bonatsos2022b}.

This paper is structured as follows: In Sec. II, a brief theoretical framework of the HFB approach and the numerical details are presented. The results and discussion of the neutron and proton-induced SC are given in Sec. III. The comparison between non-relativistic and relativistic mean-field theory predictions is considered in Sec. IV. Finally, the summary and conclusions are presented in Sec. V.
 
\section{Theoretical Framework and Numerical Details}

The Hartree-Fock-Bogoliubov theory has been extensively discussed in the literature \cite{Ring1980,Bender2003,Stoitsov2005,Stoitsov2013,Changizi2015}. In general, the HFB theory consists of the Hartree-Fock (HF) part that describes the particle-hole (p-h) channel, while the second part is the particle-particle (p-p) channel. Depending on the two channels, the density matrix ($\rho_{ij}$) and the pairing tensor density ($\kappa_{ij}$) can be used to characterize the nuclear system. The two parts were unified and treated in a general HFB formalism using a variational principle.

In the standard HFB formalism, the Hamiltonian is reduced into two parts, namely the mean field ($\Gamma$) in the particle-hole channel and the pairing field ($\Delta$) in the  particle-particle channel \cite{Stoitsov2005,Changizi2015}. It gives rise to the HFB equation \cite{Stoitsov2005}
\begin{equation}
\left(  \begin{array}{cc} e + \Gamma -\lambda & \Delta \\ -\Delta^* & -(e+\Gamma)^* +\lambda \end{array}  \right) 
\left( \begin{array}{c} U \\ V \end{array}   \right)  = E     \left(  \begin{array}{c} U \\ V \end{array}  \right), 
\end{equation}
where the Lagrange multiplier $\lambda$ has been introduced to fix the correct average particle number, $E$ denotes the energy, while the coefficients $U$ and $V$ determine the quasiparticle operator \cite{Ring1980}.  

In this work, the numerical calculations were performed using the HFBTHO code \cite{Stoitsov2013} with Skyrme-SKI3 functional \cite{Reinhard1995}. The code utilizes the axial transformed harmonic oscillator (THO) single-particle basis to expand quasi-particle wave functions. It iteratively diagonalizes the HFB Hamiltonian based on generalized Skyrme-like energy densities and a zero-range pairing interaction until a self-consistent solution is found \cite{Stoitsov2005,Stoitsov2013}. The calculations were performed with a mixed volume-surface pairing force with cut-off quasi-particle energy $E_{cut}=60$ MeV, which means that all the quasi-particles with energy lower than the cut-off are taken into account in the calculations of the densities. The code automatically sets $b_0$ by using the relation of HO frequency \cite{Stoitsov2013}
\begin{equation}
b_0 = \sqrt{\hbar \over m \omega_0},		       	
\end{equation}
 with 
 $\hbar \omega_0 =1.2 \quad 41/A^{1/3}$ (see sec. 4.2 of Ref. \cite{Stoitsov2013} for further details).		

The number of oscillator shells taken into account was $N_{max}=20$. The axial deformation parameter ($\beta_2$) of the basis taken into account was 0.300, the total number of states in the basis $N_{states}=500$, while standard HFB calculations have been performed. The maximum number of iterations in the self-consistent loop was 200, while both the direct and exchange Coulomb potentials were included. The number of Gauss-Laguerre and Gauss-Hermite quadrature points was $N_{GL}=N_{GH}=40$, and the number of Gauss-Legendre points for the integration of the Coulomb potential was $N_{Leg}=80$. The same values of the pairing strength for neutrons and protons ($V_0^{(n,p)}=-300.0$ MeV fm$^3$) have been used in the present study, given by \cite{Stoitsov2013,Changizi2015}
\begin{equation}
V_{pair}^{n,p} ({\bf r})= V_0^{n,p} \left( 1-\alpha  {\rho(\bf r) \over \rho_c} \right) \delta({\bf r}-{\bf r'}), 			    	       
\end{equation}
where $\rho({\bf r})$ is the local density, $\rho_c$ is the saturation density fixed at $\rho_c=0.16$ fm$^{-3}$, $V_0^{n,p}$ is the pairing strength, while the type of pairing force is defined by the parameter $\alpha$, which in the present study takes the value $\alpha=0.5$, related to the mixed volume-surface pairing force  \cite{Stoitsov2005,Stoitsov2013}.

It may be noticed that the decision to use the SKI3 functional in the HFBTHO code has been made after performing a set of preliminary calculations, indicating that 
SKI3 was providing better results than other Skyrme functionals, mentioned in sec. 4.2 of Ref. \cite{Stoitsov2013}. The strength of the pairing force was then set to $-300.00$ MeV fm$^3$, as described above, in order to maximize similarity of the present results to those of covariant density functional theory given in Ref. \cite{Bonatsos2022b}. Detailed comparisons to results of earlier calculations are given in Sec. IV of the present work. 

\section{Numerical results and discussion}

\subsection{Shape Coexistence due to the neutron number}

SC due to the neutron number is known \cite{Heyde2011,Bonatsos2022b} to appear around the major proton shell closures at $Z=82$ and 50, as well as around the proton subshell closure at $Z=40$. 

\subsubsection{$Z$=78-82 region}

In subsec. III.A we present the results for the particle-hole excitations (p-h), by showing the single-particle energy levels $E_{sp}$ of the proton orbitals relative to the Fermi energy $E_F$ as a function of the neutron number $N$. 

The results for the $Z$=78-82 region are shown in Figs. 1-5. These results are obtained by employing the Skyrme-SKI3 functional for the Pt, Hg, Pb, and Po isotopes using the non-relativistic self-consistent mean-field HFBTHO code \cite{Stoitsov2013}. The orbitals $1h_{11/2}$, $3s_{1/2}$, $2d_{3/2}$, $2d_{5/2}$ and $1g_{7/2}$ are forming the $Z=50$-82 proton shell, while the orbital $1h_{9/2}$ lies above $Z=82$. The orbitals are labeled by Nilsson quantum numbers $K^\pi [N n_z \Lambda]$ \cite{Nilsson1995}. The way in which Nilsson quantum numbers are assigned to the various orbitals is described in detail in sec. III.A of Ref. \cite{Schunck2010}. In addition, in order to make the observation of p-h excitations easier, we calculate the occupation probability for each orbital, shown to the right of the single-particle figures.  Occupation probability close to unity represents a fully occupied orbital (below the Fermi surface), null occupation probability represents a fully vacant orbital (above the Fermi surface), while occupation probability close to 0.5 represents a partially occupied orbital (near/at the Fermi surface).

In Fig. 1 we see that the 11/2[505] proton orbital of $1h_{11/2}$ (which should have been fully below $Z=82$) lies above the Fermi surface, therefore leaving two holes in the 50-82 shell in the regions $N= 98$-138 for Po, $N= 96$-120 for Pb, $N= 94$-114 for Hg, and $N=88$-120 for Pt isotopes. The orbital 11/2[505] is vacant in these regions.

In Fig. 2 we see that the 1/2[400] proton orbital of $3s_{1/2}$, as well as the 3/2[402] proton orbital of $2d_{3/2}$ (normally lying below $Z=82$)  lies above the Fermi surface, creating four holes in the 50-82 shell in the regions $N=98$-138 for Po, $N= 98$-120 for Pb, $N\approx 94$-114 for Hg and $N\approx  88$-120 for Pt isotopes. The 1/2[400] proton orbital of $3s_{1/2}$ and the 3/2[402] proton orbital of $2d_{3/2}$ are vacant in these regions.

In Fig. 3 we notice that the orbital $1h_{9/2}$ shows a different behavior relative to Figs. 1, 2. We see that the 1/2[541] and 3/2[532] proton orbitals of $1h_{9/2}$ (which should have been above $Z=82$) are occupied, thus creating four particle excitations in the regions $N\approx 98$-138 for Po, $N\approx 96$-120 for Pb, $N\approx  94$-114 for Hg and $N\approx  92$-120 for Pt isotopes. It is worth noting that the 1/2[541] and 3/2[532] proton orbitals of $1h_{9/2}$ approach the Fermi surface in the case of Pt isotopes, as in Ref. \cite{Bonatsos2022b}. 

From Figs. 1-3 we can identify the regions where the particle-hole excitations appear. The particle-hole excitations formed by holes can be seen in Figs. 1 and 2 in regions $N= 98$-138 for Po, $N\approx 96$-120 for Pb, $N\approx  94$-114 for Hg, and $N\approx 88$-120 for Pt, and the particle-hole excitations formed by particles are seen Fig. 3 in regions $N\approx 98$-138 for Po, $N\approx 96$-120 for Pb, $N\approx 94$-114 for Hg and $N\approx 92$-120 for Pt isotopes.

In Fig. 4  we see that all the 1/2[420], 3/2[411] and 5/2[402] proton orbitals of $2d_{5/2}$ (normally lying below $Z=82$) are occupied and remain below the Fermi surface.
Indeed, the occupation probability (right column of Fig. 4) shows that all of these orbitals lie close to unity, with the  exception of 5/2[402] in the case of the Pt isotopes in the region $N\approx 102$-114,  which starts getting above the Fermi surface, a reasonable result since Pt has fewer protons than Po, Pb, Hg, as also seen in Ref.  \cite{Bonatsos2022b}.

In a similar way in Fig. 5 we see that all the orbitals of $1g_{7/2}$ (normally lying below $Z=82$) are occupied and remain below the Fermi surface. The 7/2[404] proton orbital of $1g_{7/2}$ starts gradually approaching the Fermi surface as we move from Po to Pt isotopes, in agreement to Ref. \cite{Bonatsos2022b}.

These results will be compared in detail to earlier work in Section IV. 

It may be noticed that the study of particle-hole excitations across the Fermi level has the advantage of functioning equally well independently of the positioning of the Fermi energy near magic gaps or away from them. While neutron-induced SC usually occurs close to proton magic gaps, as seen in the present section III.A, this is not the case for proton-induced SC, which occurs away from neutron magic gaps,  as we shall see in sec. III.B.

\subsubsection{$Z$=50-52 region}

In Fig. 6, we see that the 9/2[404] orbital of $1g_{9/2}$ (which should have been below $Z=50$) is vacant and lies above the Fermi energy surface. At the same time, we see that the 3/2[422] orbital of $1g_{7/2}$ (which should have been above $Z=50$) is occupied and lies below the Fermi energy surface. As a consequence, two particles-two holes (2p-2h) proton excitations are seen in the region $N=64$-70 for Te isotopes. The proton 2h excitations are formed by the 9/2[404] orbital, while the proton 2p excitations are formed by the 3/2[422] orbital, as it is seen clearly in the occupation probability (right column of Fig. 6) of the orbitals $1g_{9/2}$ and $1g_{7/2}$.

The present results with Skyrme-SKI3 functional are in agreement with the results of Ref. \cite{Bonatsos2022b} for covariant density functional theory with the DDME2 functional in the region $N=64$-68, with the exception of $N=70$, which appears in our study, while it is not found in Ref. \cite{Bonatsos2022b}. Also they are in agreement with the results obtained by a relativistic Hartree-Bogoliubov approach for the Te isotopes in the region $N=64$-68 \cite{Sharma2019}.

\subsubsection{$Z$=38 and 40 region}

In Fig. 7, we see that the 1/2[301] orbital of $2p_{1/2}$, as well as the 3/2[301] and 5/2[303] orbitals of $1f_{5/2}$ (which should lie below $Z=40$) are vacant and lie above the Fermi energy surface. At the same time, we see that the 1/2[440], 3/2[431] and 5/2[422] orbitals of $1g_{9/2}$ (which should be above $Z=40$) are occupied and lie below the Fermi energy surface. As a consequence, 6p-6h proton excitations are seen in the region $N=36$-42 for Zr ($Z=40$) isotopes. The proton 6h excitations are formed by the 1/2[301], 3/2[301], and 5/2[303] orbitals, while the proton 6p excitations  are formed by the 1/2[440], 3/2[431] and 5/2[422] orbitals in the regions $N=36$-42 for Zr isotopes. This is seen clearly in the occupation probabilities (right column of Fig. 7) of the orbitals $1f_{5/2}$,  $2p_{1/2}$ and $1g_{9/2}$.

The present results are in agreement with the results of Ref. \cite{Bonatsos2022b} for covariant density functional theory with the DDME2 functional in regions $N=36$-42, with the exception of $N=36$ and 42, which appear in our study, while they are not found in Ref. \cite{Bonatsos2022b}. Moreover, the present results are in agreement with the results of beyond mean field methods using the Gogny D1S interaction in the region $N=40$ \cite{Rodriguez2011}, as well as with the results of the total-Ruthian-surface calculations \cite{Zheng2014} in the regions $N=38$-42.

In Fig. 8, we see that the 3/2[301] and 5/2[303] orbitals of $1f_{5/2}$ (which should lie below $Z=40$) are vacant and lie above the Fermi energy surface, while the 1/2[440] and 3/2[431] orbitals of $1g_{9/2}$ (which should lie above $Z=40$) are occupied and lie below the Fermi energy surface. As a consequence, 4p-4h proton excitations are seen in the region $N=46$ for Sr isotopes. The proton 4h excitations are formed by the 1/2[440] and 3/2[431] orbitals, while the proton 4p excitations are formed by the 3/2[301] and 5/2[303] orbitals in the region $N=46$ for Sr isotopes. This is seen in the occupation probabilities (right column of Fig. 8) of the orbitals $1f_{5/2}$ and $1g_{9/2}$. 
 
Based on the occupation probability of proton orbitals $1f_{5/2}$,  $2p_{1/2}$ and $1g_{9/2}$ for Sr ($Z=38$) isotopes (right column of Fig. 8), the present results show an island of SC in the region of $N=46$, while the results of Ref. \cite{Bonatsos2022b} for covariant density functional theory with the DDME2 functional show an island of SC in the region of $N=40$. In addition, the results of the total-Ruthian-surface \cite{Zheng2014} show an island of SC in the region $N=38$-42. 

The results of the present section are summarized in Table I.

\subsection{Shape Coexistence due to the proton number}

SC due to the proton number is known \cite{Bonatsos2022a,Bonatsos2022b} to appear around the neutron numbers $N=90$, 60, and 40 \cite{Heyde2011}, corresponding to p-h excitations across the harmonic oscillator magic numbers 112, 70, and 40, due to the beginning of the occupation of the intruder orbitals $1i_{13/2}$, $1h_{11/2}$, and $1g_{9/2}$ respectively.    

\subsubsection{$N$=90-96 region}

In subsec. III.B we present the results for the particle-hole excitations (p-h), by showing the single-particle energy levels $E_{sp}$ of the neutron orbitals relative to the Fermi energy $E_F$ as a function of the proton number $Z$.

The results for the $N=90$-96 region are shown in Figs. 9-11. 
The results are obtained by employing the Skyrme-SKI3 functional for the $N=90$-96 isotones using the non-relativistic self-consistent mean-field HFBTHO code of Ref. \cite{Stoitsov2013}. The orbitals $1i_{13/2}$, $2f_{7/2}$ and $1h_{9/2}$ are considered. In addition, in order to make the observation of p-h excitations easier, we calculate the occupation probability for each orbital, shown to the right of the single-particle figures. 
 
In Fig. 9, we see that the 1/2[660] neutron orbital of $1i_{13/2}$ (which should have been above $N=112$) is occupied and lies below the Fermi surface in the case of the $N=90$ isotones for $Z=58$-68, thus becoming a candidate for being the particle partner in 2p-2h neutron excitations. In the case of $N=92$ isotones, we see that both the 1/2[660] and 3/2[651] neutron orbitals of $1i_{13/2}$ are sunk below the Fermi surface for $Z=58$-68, thus becoming  candidates for being the particle partners in 4p-4h neutron excitations. The same is happening in the cases of $N=94$ and 96 neutrons for $Z=58$-66. 

In Fig. 10, we see that the 5/2[523] neutron orbital of $2f_{7/2}$ (which should have been below $N=112$) is empty and lies above the Fermi surface for the case of $N=90$ and 92 isotones for $Z=58$-68, thus becoming a candidate for being the hole partner in 2p-2h neutron excitations, while it reaches the Fermi surface for $Z= 58$-66 (with occupation probability 0.15-0.44, respectively) in the case of the $N=94$ isotones and sinks below the Fermi surface in the case of the $N=96$ isotones. 

In Fig. 11, we see that the 3/2[521] neutron orbital of $1h_{9/2}$ is empty and lies above the Fermi surface for the case of $N=90$ and 92 isotones at $Z= 58$-68, thus becoming a candidate for being the hole partner in 2p-2h excitations, while in the case of the $N=94$ isotones it is sinking lower and lower in relation to the Fermi surface at $Z= 58$-68, 74-78 with partial occupation probability $\cong 0.5$ (as seen in the right column of Fig. 11), while it falls completely below the Fermi surface with occupation probability 
$\cong 0$ in the case of the $N=96$ isotones . 

From Figs. 10, 11 we see that particle-hole excitations formed by holes arise in the $N=90$, 92 isotones for the 5/2[523] neutron orbital of $2f_{7/2}$ and the 3/2[521] neutron orbital of $1h_{9/2}$ at $Z=58$-68, while in the $N=94$ isotones they arise in the 5/2[523] neutron orbital of $2f_{7/2}$ at $Z=58$-66.

From Fig. 9 we see that particle-hole excitations formed by particles arise in the $N=90$ isotones for the 1/2[660] neutron orbital of $1i_{13/2}$ at $Z=58$-68,  in the $N=92$ isotones  for  the 1/2[660] and 3/2[651] neutron orbitals of $1i_{13/2}$ at $Z=58$-68 , while in the $N=94$ isotones they arise for the 1/2[660] and 3/2[651] neutron orbitals of $1i_{13/2}$ at $Z=58$-66.

The present results in Figs. 9-11 are in agreement in the case of $N=90$, 92 isotones at $Z=60$-64 with the results of covariant density functional theory with the DDME2 functional \cite{Bonatsos2022b} and relativistic energy density functionals \cite{Li2009}, with the exception of some regions which are found in our study with the Skyrme-SKl3 fuctional but they are not found in Ref. \cite{Bonatsos2022b}, which are the $N=90$, 92 isotones at $Z=58$, 66, 68, as well as  the $N=94$ isotones in the region $Z=58$-66. 

In Fig. 9 we observe that the 1/2[660] orbital, while being below the Fermi energy at middle values of $Z$, jumps well above it both at low and high values of $Z$, crossing the other orbitals, while, in contrast, in Fig. 10 the lowest-lying 1/2[541] orbital seems to converge to the other orbitals at high values of $Z$. In Figs. 10 and 11 of Ref. \cite{Bonatsos2022b}, in which the same orbitals are shown in the framework of covariant density functional theory with the DDME2 functional, convergence is seen in both cases. This unusual crossing is also observed, though to a much lesser extent, in Figs. 7 and 8 of the present work for the 1/2[440] orbital, again differing from Figs. 8 and 9 of Ref. \cite{Bonatsos2022b}, in which convergence is seen. In any case, these unusual crossings of single-particle orbitals, which might be attributed to the way in which Nilsson quantum numbers are assigned to the various levels within the two different approaches (see sec. III.A of Ref. \cite{Schunck2010} for the present Skyrme HFB approach and sec. I of Ref. \cite{Bonatsos2022b}, as well as Refs. \cite{Prassa2013,Karakatsanis2017,Karakatsanis2020} for the covariant density functional theory case),  do not affect the conclusions of the present study.

\subsubsection{$N$=60 region}

Fig. 12 shows the 4p-4h proton excitations in the region of the $N=60$ isotones. We see that the 1/2[411] orbital of $3s_{1/2}$ and the 5/2[413] orbital of $2d_{5/2}$ (which should have been below $N=70$) are empty, thus creating 4h proton excitations at $Z=36$-44, while the 1/2[550] and 3/2[541] orbitals of $1h_{11/2}$ (which should have been above $N=70$) are occupied, thus creating 4p proton excitations at $Z=36$-44.

The present results are in agreement with the results of Ref. \cite{Bonatsos2022b} for covariant density functional theory with the DDME2 functional in the region of the $N=60$ isotones at $Z=40$ and 42, with the exception of $Z=36$, 38 and 44,  which appear in our study, while they are not found in Ref. \cite{Bonatsos2022b}. In addition, the present results are in agreement with the results for $N=60$ of configuration mixing calculations \cite{Wu2003} for Sr and Zr and IBM-CM calculations for Zr \cite{GarciaRamos2020}, as well as with experimental SC in $N=60$ isotones at $Z=42$ ($_{42}^{102}$Mo$_{60}$) \cite{Esmaylzadeh2021}.

\subsubsection{$N$=58 region}

Fig. 13 shows the 4p-4h proton excitations in the region of the $N=58$ isotones. We see that the 1/2[411] orbital of $3s_{1/2}$ and the 5/2[413] orbital of $2d_{5/2}$ (which should have been below $N=70$) are empty everywhere, thus creating 4h proton excitations at $Z=36$-42, while the 1/2[550] and 3/2[541] orbitals of $1h_{11/2}$ (which should have been above $N=70$) are occupied, thus creating 4p proton excitations at $Z=36$-42.

The present results are in agreement with the results of Ref. \cite{Bonatsos2022b} for CDFT with the DDME2 functional in the region of the $N=58$ isotones at $Z=40$, with the exception of $Z=36$, 38 and 42, which appear in our study, while they are not found in Ref. \cite{Bonatsos2022b}. In addition, the present results are in agreement with the results of the VAMPIR approach \cite{Petrovici2012} in the region of $N=58$ isotones at $Z=40$ ($_{40}^{98}$Zr$_{58}$).

\subsubsection{$N$=38 and 40 region}

Fig. 14 shows the 6p-6h proton excitations in the region of $N=40$ isotones. We see that the 3/2[301] and 5/2[303] orbitals of $1f_{5/2}$ and the 1/2[301] orbital of $2p_{1/2}$ (which should have been below $N=40$) are empty, thus creating 6h proton excitations at $Z=36$-42, while the 3/2[431], 1/2[440] and 5/2[422] orbitals of $1g_{9/2}$ (which should have been above $N=40$) are occupied, thus creating 6p proton excitations at $Z=36$-42. The orbital 9/2[404] of $1g_{9/2}$, not shown in Fig. 14, is the highest-lying orbital in the 28-50 shell, thus it is not occupied. 

The present results are in agreement with the results of Ref. \cite{Bonatsos2022b} for CDFT with the DDME2 functional in the region of $N=40$ isotones at $Z=40$, with the exception of $Z=36$, 38 and 42, which  appear in our study, while they are not found in Ref. \cite{Bonatsos2022b}.

Fig. 15 shows the 4p-4h proton excitations in the region of $N=38$ isotones. We see that the 3/2[301] and 5/2[303] orbitals of $1f_{5/2}$ (which should have been below $N=40$) are empty, thus creating 4h proton excitations at $Z=32$-42, while the 3/2[431], 1/2[440] orbitals of $1g_{9/2}$ (which should have been above $N=40$) are occupied, thus creating 4p proton excitations at $Z=32$-42. The orbital 9/2[404] of $1g_{9/2}$, not shown in Fig. 15, is the highest-lying orbital in the 28-50 shell, thus it is not occupied. 

The present results are in agreement with the results of Ref. \cite{Bonatsos2022b} for CDFT with the DDME2 functional in the region of $N=38$ isotones at $Z=40$, with the exception of Z=32, 34, 36, 38 and 42, which appear in our study, while they are not found in Ref. \cite{Bonatsos2022b}.

The results of the present section are summarized in Table II. 

\section{Comparison to earlier results in the ${\bf Z=78}$-82 region} 

The islands of shape coexistence found in both subsecs. III.A and III.B using non-relativistic self-consistent mean-field HFB theory with Skyrme-SKI3 functional are shown in Fig. 16, in which they are compared with the results of Ref. \cite{Bonatsos2022b}, obtained with covariant density functional theory with the DDME2 functional.

In Fig.  16 we remark that our predictions for shape coexistence (SC) lie within the regions predicted by the proxy-SU(3) symmetry, 
with notable exceptions occurring for the neutron-rich Po isotopes with $N=114$-138, Pb and Pt isotopes with $N=114$-120, as well as for the Hg isotope with $N=114$. 
In addition, some exceptions occur for the neutron-deficient Pt isotopes with $N=88$-94 and the Hg isotope with $N=94$. It is therefore of interest to examine 
to what extent these predictions are supported by alternative theoretical approaches and/or experimental data. In the following subsections, only studies 
regarding isotopes above $N=114$ or below $N=94$ will be considered. 

\subsection{The Po isotopes} 

Energy systematics for low-lying states for the Po isotopes with $N=110$-126 are given in Fig. 4 of Ref. \cite{Younes1997}, as well as in Fig. 1 of \cite{Kesteloot2015}. The parabola formed by the $0_2^+$ states, which is a landmark of SC, is seen up to $N=118$. However, beyond $N=116$ the excitation energy of the $0_2^+$ state exceeds 1 MeV, thus one cannot speak about SC of the ground state band and a band built on the $0_2^+$ state, with SC thus limited to $N=114$ and below, in agreement to the results obtained in \cite{Xu2007} within a density-dependent cluster model. This is in agreement also with the suggestion of existence of 4p-2h proton excitations across the $Z=82$ shell closure in the Po isotopes with $N=112$, 114, similar to the 2p-2h proton excitations known to occur in the Pb isotopes, as suggested in \cite{Alber1991}. The existence of 4p-2h excitations also for the Po isotope with $N=116$ has been suggested in \cite{Bijnens1998}, while for $N=118$ a picture with two-phonon multiplets in a quadrupole vibrational model have been suggested \cite{Bijnens1998}. 

Early calculations within the extended Interacting Boson Model (EIBM), which takes into account particle-hole excitations, have been successfully performed for Po isotopes up to $N=116$ \cite{DeCoster1999}, while more recent Interacting Boson Model with configuration mixing (IBM-CM) calculations have been successfully performed in Po isotopes up to $N=124$ \cite{GarciaRamos2015}, indicating a gradual transition from configuration mixed states to pure regular states at different neutron number $N$, depending on the angular momentum of the states. In particular, the $0_2^+$ state appears to be purely intruder up to $N=116$, becoming pure regular at $N=120$. The $2_1^+$ and $4_1^+$ states become regular at $N=116$, while the $2_2^+$ and $4_2^+$ states become regular at $N=118$. As a consequence, the SC phenomenon becomes hidden beyond $N=116$, the situation resembling the one appearing in the Pt isotopes
\cite{GarciaRamos2015}.       

Similar conclusions have been reached in a study of the Po isotopes with $N=108$-126 using a simple two-state mixing approach, as well as an IBM-2 approach \cite{Oros1999}. Good results have been obtained for the Po isotopes with $N=108$-114, while for the Po isotopes with $N=116$-126 the need for further efforts has been pointed out \cite{Oros1999}. 
Configuration-constrained potential-energy-surface calculations \cite{Shi2010}, performed within a macroscopic-microscopic model using a deformed Woods-Saxon potential and the Lipkin-Nogami pairing have also obtained a good description of the Po isotopes up to $N=116$. 

An extended study of the Po isotopes with $N=102$-120 has been performed within a beyond mean-field  calculation using a Skyrme energy density functional \cite{Yao2013}, finding a change of structure for the Po isotopes occurring at $N=114$, with the onset of oblate deformation already taking place at $N=112$ (see also \cite{WrzosekLipska2016}). 

In summary, existing work seems to support SC at most up to $N=116$ and not beyond this point. 

\subsection{The Pb isotopes} 

2p-2h proton excitations in the Pb isotopes have been identified experimentally for $N=110$-118 \cite{VanDuppen1984,VanDuppen1987}, but collective bands built on the $0_2^+$ state, meaning the occurrence of shape coexistence, have been identified only up to $N=114$ \cite{VanDuppen1987,Penninga1987}. 

It has been shown that 2p-2h proton excitations are expected to occur in a wide range of Pb isotopes \cite{Heyde1989}, but applications of the EIBM \cite{DeCoster1997} and the IBM-CM \cite{Hellemans2008}, related to configuration mixing and SC, have only been extended up to $N=114$ \cite{DeCoster1997,Hellemans2008}. In parallel, configuration-constrained potential-energy-surface calculations \cite{Shi2010}, performed within a macroscopic-microscopic model using a deformed Woods-Saxon potential and the Lipkin-Nogami pairing have also obtained a good description of the Pb isotopes up to $N=116$, as in the Po case \cite{Shi2010}. 

In summary, existing work seems to support SC at most up to $N=114$ and not beyond this point.

A long discussion has been made \cite{Yoshida1994,Heyde1996,Takigawa1996} on the nature of the ground states of the Pb isotopes with $N=106$-114, predicted to have oblate deformed shapes within a deformed relativistic mean field (RMF) calculation \cite{Yoshida1994}, in contrast to a large body of experimental evidence pointing to spherical shapes for these ground states \cite{Heyde1996}. The consensus reached \cite{Heyde1996,Takigawa1996} indicates that the ordering between the prolate and oblate orbitals is very sensitive to the input parameters in the RMF Lagrangian, as well as to the choice of the pairing gap \cite{Takigawa1996}.   

\subsection{The Hg isotopes} 

Recent empirical evidence \cite{Olaizola2019} suggests minimal mixing between normal and intruder structures for the Hg nuclei with $N=112$-120, thus limiting the region presenting SC up to $N=110$, in agreement to early Nilsson-Strutinsky calculations \cite{Nazarewicz1993}, as well as to more recent IBM-CM calculations \cite{GarciaRamos2014a} extended up to $N=120$. 
Recent IBM-2 calculations with parameters microscopically determined through mean-field calculations using the Gogny energy density functional, extended up to $N=124$, also find SC 
up $N=110$ \cite{Nomura2013}.  

In summary, existing work seems to support SC at most up to $N=110$ and not beyond this point.

A long discussion has been made \cite{Patra1994,Heyde1996,Takigawa1996} on the nature of the ground states of the Hg isotopes with $N=100$-110, for which a large body of experimental evidence points to oblate shapes for their ground states and to prolate shapes for their excited bands \cite{Heyde1996}, while a deformed relativistic mean field (RMF) calculation \cite{Patra1994} predicted prolate ground states for $N=98$-108. The consensus reached \cite{Heyde1996,Takigawa1996} indicates that the ordering between the prolate and oblate orbitals is very sensitive to the input parameters in the RMF Lagrangian, as well as to the choice of the pairing gap \cite{Takigawa1996}. More recent mean-field calculations using a Skyrme energy density functional \cite{Yao2013} suggest the existence of oblate ground states in the Hg isotopes above $N=106$ and prolate ground states below it. It becomes evident that the details of the interaction largely affect the relative ordering of prolate and oblate orbitals, as in the case of the Pb isotopes considered in the previous subsection.  

\subsection{The Pt isotopes} 

Early calculations \cite{Bengtsson1987} using the Strutinsky renormalization plus BCS pairing approach, and involving the Woods-Saxon potential or the modified harmonic oscillator (Nilsson) potential, find secondary minima corresponding to excited $0^+$ states coexisting with the ground state for the Pt isotopes with $N=98$-114.  

Calculations using the IBM-CM approach find configuration mixing in the Pt isotopes up to $N=116$ and down to $N=94$ \cite{Harder1997,GarciaRamos2011}, arguing in addition that clear signs of shape coexistence are seen between $N=98$ and $N=108$, where the intruder state becomes the ground state \cite{GarciaRamos2014b}. It is however remarked \cite{GarciaRamos2011} that the presence of two different structures in the Pt isotopes is not as clear in the Pt isotopes as in the Hg isotopes \cite{GarciaRamos2014a}, rather resembling the situation in the Po isotopes \cite{GarciaRamos2015}, in which the SC phenomenon is rather  hidden. These observations are in agreement with the interpretation of the properties of the Pt isotopes with $N=94$-116 in terms of an IBM Hamiltonian without intruder states and/or  configuration mixing \cite{McCutchan2005}.    

A recent calculation \cite{Majarshin2021} 	involving a pairing Hamiltonian with configuration mixing for the $N=116$ isotope of Pt shows that the mixing is small, while the ground state band is only slightly more collective than the excited-state band, thus providing no evidence for SC. 

Extensive calculations \cite{Nomura2011b} involving an IBM Hamiltonian with parameters determined from Hartree-Fock-Bogoliubov (HFB) calculations with the microscopic Gogny energy density functional D1M  find that in the Pt isotopes with $N=110$-122 the transition from prolate to oblate shapes occurs much more slowly than in the Yb, Hf, W, and Os series of isotopes. Similar results are obtained for the Er-Pt series of isotopes with a five-dimensional collective Hamiltonian (5DCH) based on the PC-PK1 energy density functional \cite{Yang2021a}, finding a rapid prolate to oblate transition, related to SC, at $N=116$ in the  Er and Yb series of isotopes. 

In the non-relativistic mean-field framework, extensive self-consistent axially-deformed Hartree-Fock calculations \cite{Stevenson2005} have been performed for the Yb-Pt series of isotopes with $N=110$-122, finding coexistence of prolate and oblate shapes at $N=116$, with weak dependence on the proton number. These results have been corroborated through self-consistent axially-symmetric Skyrme Hartree-Fock plus BCS calculations \cite{Sarriguren2008}, also pointing to a transition from prolate to oblate shapes at $N=116$-118, with the energies of the prolate and oblate shapes being nearly degenerate. 

In summary, there are arguments favoring the presence of SC in the region $N=96$-116, but there are also opposing views, focusing on the transition from prolate to oblate shapes 
around $N=116$ and arguing that SC is seen only around this neutron number.  

\subsection{Conclusion}

In summary, according to the literature considered in the previous subsections, SC in the Po, Pb, and Hg isotopes seems to appear up to $N=116$, 114, 110 respectively, in qualitative agreement to the present results shown in Fig. 16, which do exhibit a decrease of the maximum value of $N$ in which SC appears with decreasing $Z$ in these series of isotopes. The comeback of the limiting value for appearance of SC to $N=116$ for the Pt isotopes, suggested in the literature considered, is also seen in Fig. 16.  

In conclusion, there is some evidence for SC at $N=114$, 116 in the Pt, Hg, Pb, and Po isotopes, but practically no evidence for SC far above $N=116$, as predicted 
in the present approach for the Po isotopes. It seems that the sensitivity of the ordering of the prolate and oblate orbitals on the details of the mean field and pairing interactions  used, must be responsible for this discrepancy, calling for further comparative studies of mean-field methods in this region of the nuclear chart. 

\section{Summary and conclusions}

In this article, the islands of shape coexistence for $Z=38$ to 84 have been studied in the framework of non-relativistic self-consistent mean-field HFB theory using the Skyrme-SKI3 functional, using the particle-hole excitations mechanism in the proton and neutron single-particle energy levels relative to the Fermi energy.   
The results of the present study are compared with those obtained in the framework of covariant density functional theory with the DDME2 functional, using the same particle-hole excitations mechanism. The islands of SC that appeared in CDFT with the DDME2 functional are also appearing in the present study with the Skyrme-SKI3 functional, thus confirming the robustness of the particle-hole excitations mechanism in searching for islands of SC. In addition, the current study revealed new regions of SC, bordering the previous ones, thus increasing the total area of the islands.  

It would be interesting to perform further relativistic or non-relativistic mean-field theory calculations, employing different energy density functionals, as well as different strengths of the pairing force, in order to check the robustness of the results presented in this study and clarify the physical reasons behind any differences in the shores of the predicted islands of SC. 


\begin{turnpage}


\begin{table*}  

\caption{Proton single-particle energy levels participating in proton particle-hole formation in different regions of the nuclear chart. Since the neutrons act as elevators 
for the creation of the proton p-h excitations, we say that we have neutron-induced SC. The results obtained in this study using non-relativistic mean-field HFB theory with Skyrme-SKI3 functional (labeled by Skyrme-SKI3) are compared with the results of Ref. \cite{Bonatsos2022b} for covariant density functional theory with the DDME2 functional for the same mechanism (labeled by DDME2). The new cases of SC are summarized in the last column. See Section III.A for further discussion.
}

\bigskip

\begin{tabular}{ l  c  c c c l}

\hline
nuclei                           & occupied $Z>82$ & occupied $Z>82$ & vacant $Z<82$ & vacant $Z<82$ &  new cases of SC \\ 
                                 &    Skyrme-SKI3 &  DDME2           &  Skyrme-SKI3  &  DDME2        &            \\
\hline 
\medskip

$Z$=78, $N$=88-120 & 1/2[541] 3/2[532]& 1/2[541] 3/2[532]& 11/2[505] 1/2[400] 3/2[402]& 1/2[400] 3/2[402]& $Z$=78, $N$=88-96, 112-120 \\ \medskip
 
$Z$=80, $N$=94-114 & 1/2[541] 3/2[532]& 1/2[541] 3/2[532]& 11/2[505] 1/2[400] 3/2[402]& 11/2[505] 1/2[400] 3/2[402]& $Z$=80, $N$=94,112,114 \\ \medskip

$Z$=82, $N$=96-120 & 1/2[541] 3/2[532]& 1/2[541] 3/2[532]& 11/2[505] 1/2[400] 3/2[402]& 11/2[505] 1/2[400] 3/2[402]& $Z$=82, $N$=96,110-120 \\ \medskip 
 
$Z$=84, $N$=98-138 & 1/2[541] 3/2[532]& 1/2[541] 3/2[532]& 11/2[505] 1/2[400] 3/2[402]& 11/2[505] 1/2[400]& $Z$=84, $N$=110-138           \\ 
\hline

\hline
nuclei                         & occupied $Z>50$  & occupied $Z>50$  & vacant $Z<50$              & vacant $Z<50$ &  new cases of SC \\ 
                               &    Skyrme-SKI3   &  DDME2           &  Skyrme-SKI3               &  DDME2        &                  \\
\hline 
\medskip
 
$Z$=52, $N$=64-70  &     3/2[422]     &       3/2[422]   &   9/2[404]                 &   9/2[404]    &   $Z$=52, $N$=70  \\ 

\hline
nuclei                         & occupied $Z>40$  & occupied $Z>40$  & vacant $Z<40$              & vacant $Z<40$ &  new cases of SC \\ 
                               &    Skyrme-SKI3   &  DDME2           &  Skyrme-SKI3               &  DDME2        &                  \\
\hline 
\medskip

$Z$=38, $N$=46          & 1/2[440] 3/2[431]& 1/2[440] 3/2[431]& 3/2[301] 5/2[303]          & 3/2[301] 5/2[303] &  $Z$=38, $N$=46  \\  \medskip

$Z$=40, $N$=36-42    & 1/2[440] 3/2[431] 5/2[422]&  1/2[440] 3/2[431] 5/2[422]& 1/2[301]  3/2[301] 5/2[303] &  1/2[301]  3/2[301] 5/2[303] &  $Z$=40, $N$=36,42 \\ 

\hline
\end{tabular}
\end{table*}


\begin{table*} 

\caption{Neutron single-particle energy levels participating in neutron particle-hole formation in different regions of the nuclear chart. Since the protons act as elevators 
for the creation of the neutron p-h excitations, we say that we have proton-induced SC. The results obtained in this study using non-relativistic mean-field HFB theory with Skyrme-SKI3 functional (labeled by Skyrme-SKI3) are compared with the results of Ref. \cite{Bonatsos2022b} for covariant density functional theory with the DDME2 functional for the same mechanism (labeled by DDME2). The new cases of SC are summarized in the last column. See Section III.B for further discussion.
}

\begin{tabular}{ l  c  c c c l}

\hline
nuclei                           & occupied $N>112$ & occupied $N>112$ & vacant $N<112$ & vacant $N<112$ &  new cases of SC \\ 
                                 &    Skyrme-SKI3 &  DDME2           &  Skyrme-SKI3  &  DDME2        &            \\
\hline 
\medskip

$Z$=58-68, $N$=90 & 1/2[660]          & 1/2[660] & 5/2[523] 3/2[521] & 5/2[523] & $Z$=58,66,68, $N$=90 \\ \medskip 

$Z$=58-68, $N$=92 & 1/2[660] 3/2[651] & 3/2[651] & 5/2[523] 3/2[521] & 5/2[523] & $Z$=58,66,68, $N$=92 \\ \medskip  

$Z$=58-66, $N$=94 & 1/2[660] 3/2[651] &          & 5/2[523] 3/2[521] &          & $Z$=58-66, $N$=94 \\
\hline

\hline
nuclei                           & occupied $N>70$ & occupied $N>70$ & vacant $N<70$ & vacant $N<70$ &  new cases of SC \\ 
                                 &    Skyrme-SKI3 &  DDME2           &  Skyrme-SKI3  &  DDME2        &            \\
\hline 
\medskip

$Z$=36-42, $N$=58  & 1/2[550] 3/2[541] & 1/2[550]          & 1/2[411] 5/2[413] & 1/2[411] 5/2[413] & $Z$=36,38,42, $N$=58 \\ \medskip 

$Z$=36-44, $N$=60  & 1/2[550] 3/2[541] & 1/2[550] 3/2[541] & 1/2[411] 5/2[413] & 1/2[411] 5/2[413] & $Z$=36,38,44, $N$=60 \\  
\hline

\hline
nuclei                           & occupied $N>40$ & occupied $N>40$ & vacant $N<40$ & vacant $N<40$ &  new cases of SC \\ 
                                 &    Skyrme-SKI3 &  DDME2           &  Skyrme-SKI3  &  DDME2        &            \\
\hline 
\medskip

$Z$=32-42, $N$=38 & 1/2[440] 3/2[431]          & 1/2[440] 3/2[431]          & 3/2[301] 5/2[303]           & 3/2[301] 5/2[303]           & $Z$=32-38,42, $N$=38 \\ \medskip 

$Z$=36-42, $N$=40   &  1/2[440] 3/2[431] 5/2[422] & 1/2[440] 3/2[431] 5/2[422] & 1/2[301]  3/2[301] 5/2[303] & 1/2[301]  3/2[301] 5/2[303] & $Z$=36,42, $N$=40   \\                  
\hline

\hline
\end{tabular}

\end{table*}

\end{turnpage}


\begin{figure*} [htb]

    \includegraphics[width=175mm]{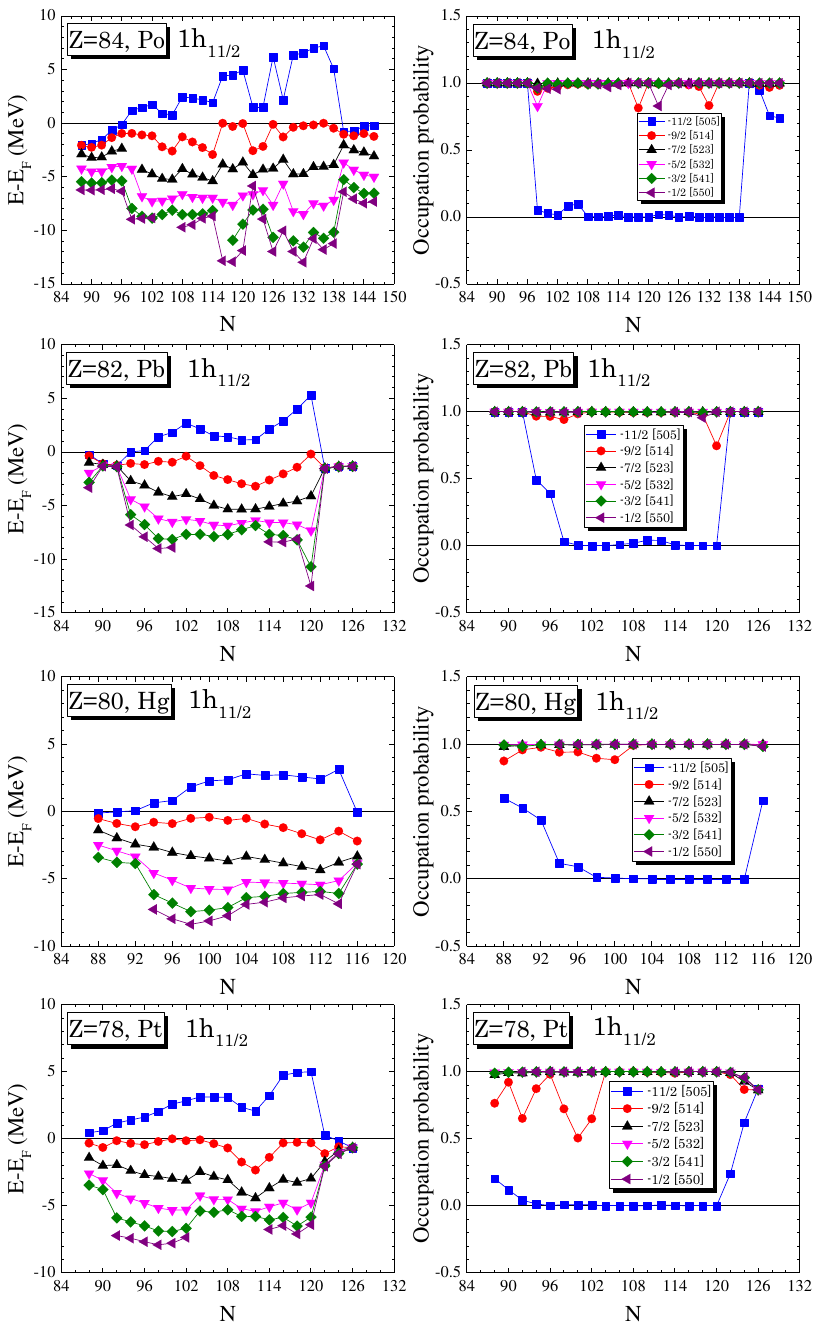}
     
    \caption{(Color online) Left column: single-particle energies of the proton orbitals of $1h_{11/2}$ relative to the Fermi energy as a function of the neutron number $N$ obtained by non-relativistic mean-field HFB theory with Skyrme-SKI3 functional for $Z=78$-84 isotopes. Right column: occupation probability of the same proton orbitals. See Sec. III.A.1 for further discussion.}
        
\end{figure*}


\begin{figure*} [htb]

    \includegraphics[width=175mm]{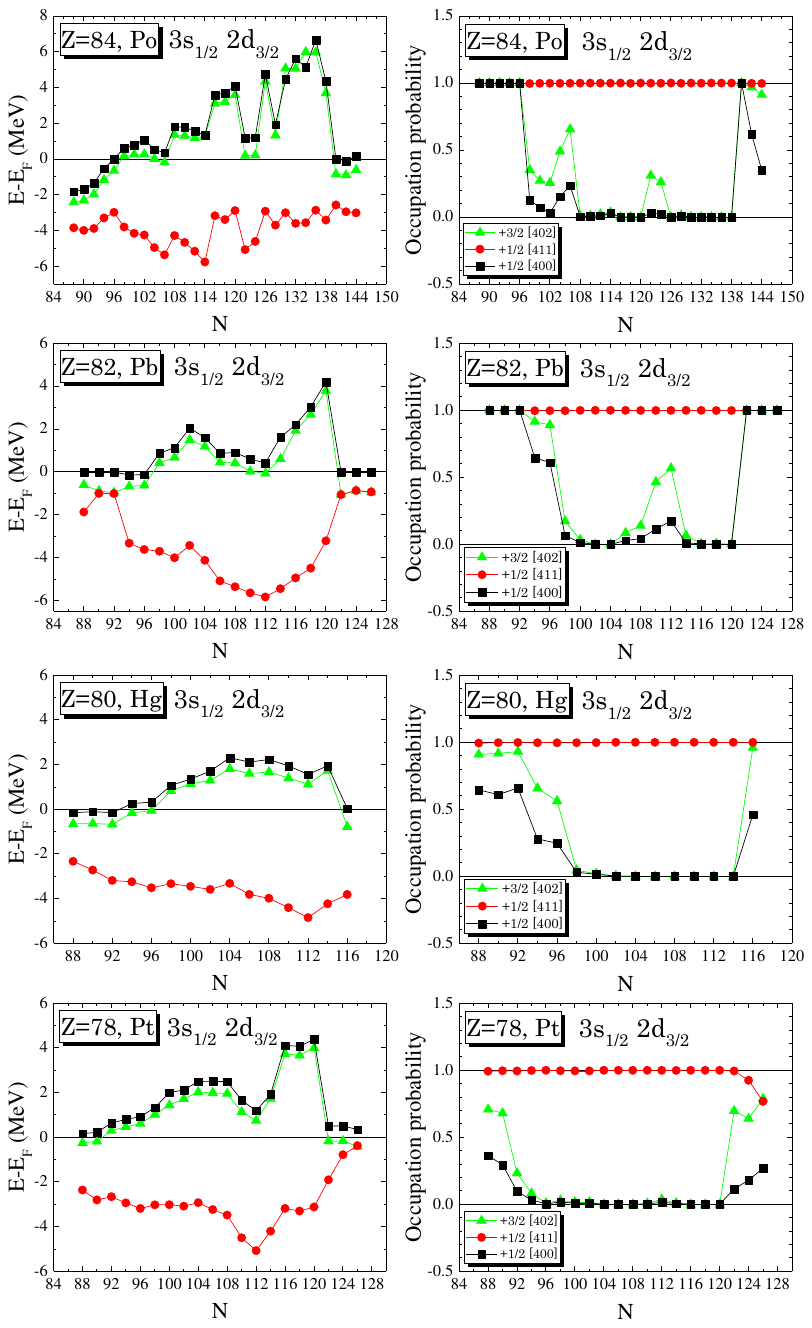}
     
    \caption{(Color online) Same as Fig. 1, but for the proton orbitals $3s_{1/2}$ and $2d_{3/2}$. See Sec. III.A.1 for further discussion.}
   
\end{figure*}


\begin{figure*} [htb]

    \includegraphics[width=175mm]{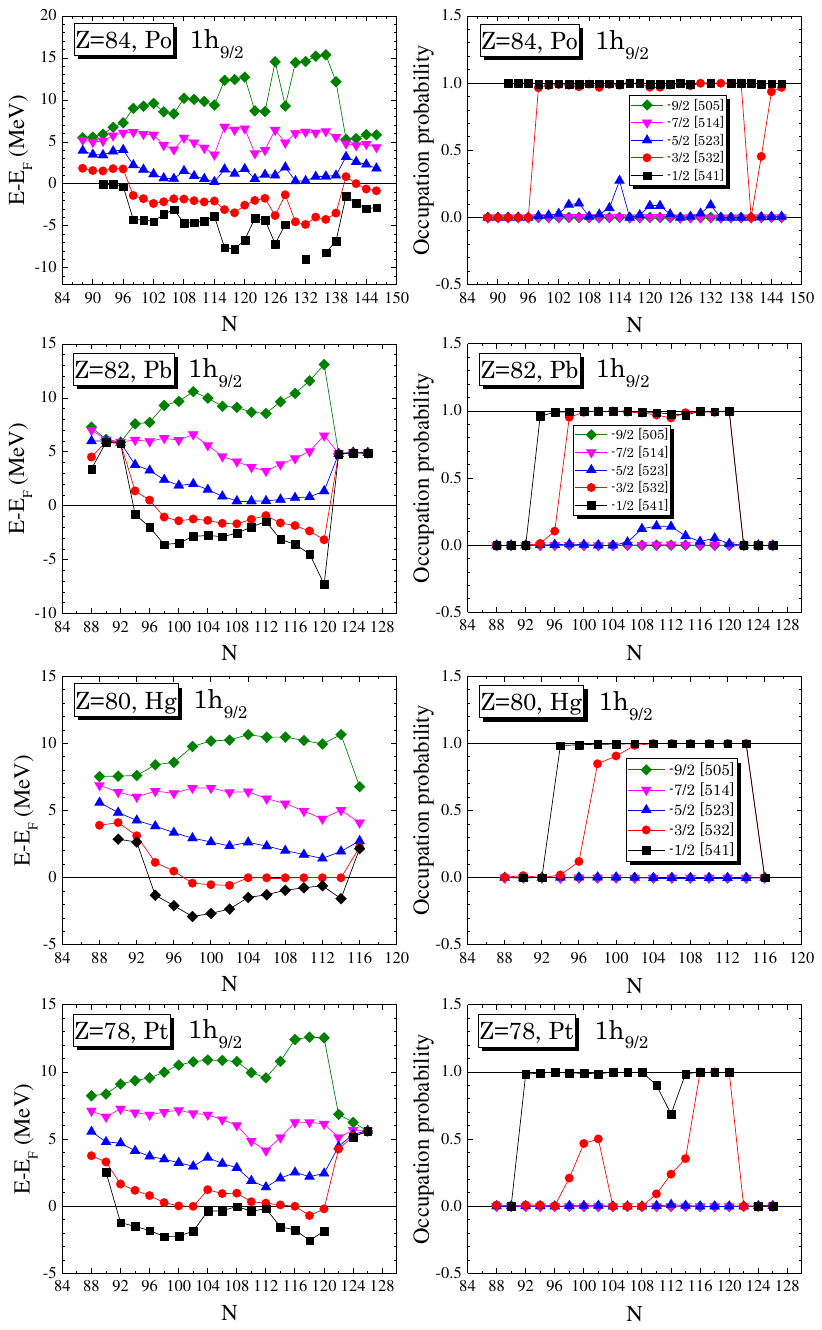}
     
    \caption{(Color online) Same as Fig. 1, but for the proton orbital $1h_{9/2}$. See Sec. III.A.1 for further discussion.}
    
\end{figure*}


\begin{figure*} [htb]

    \includegraphics[width=175mm]{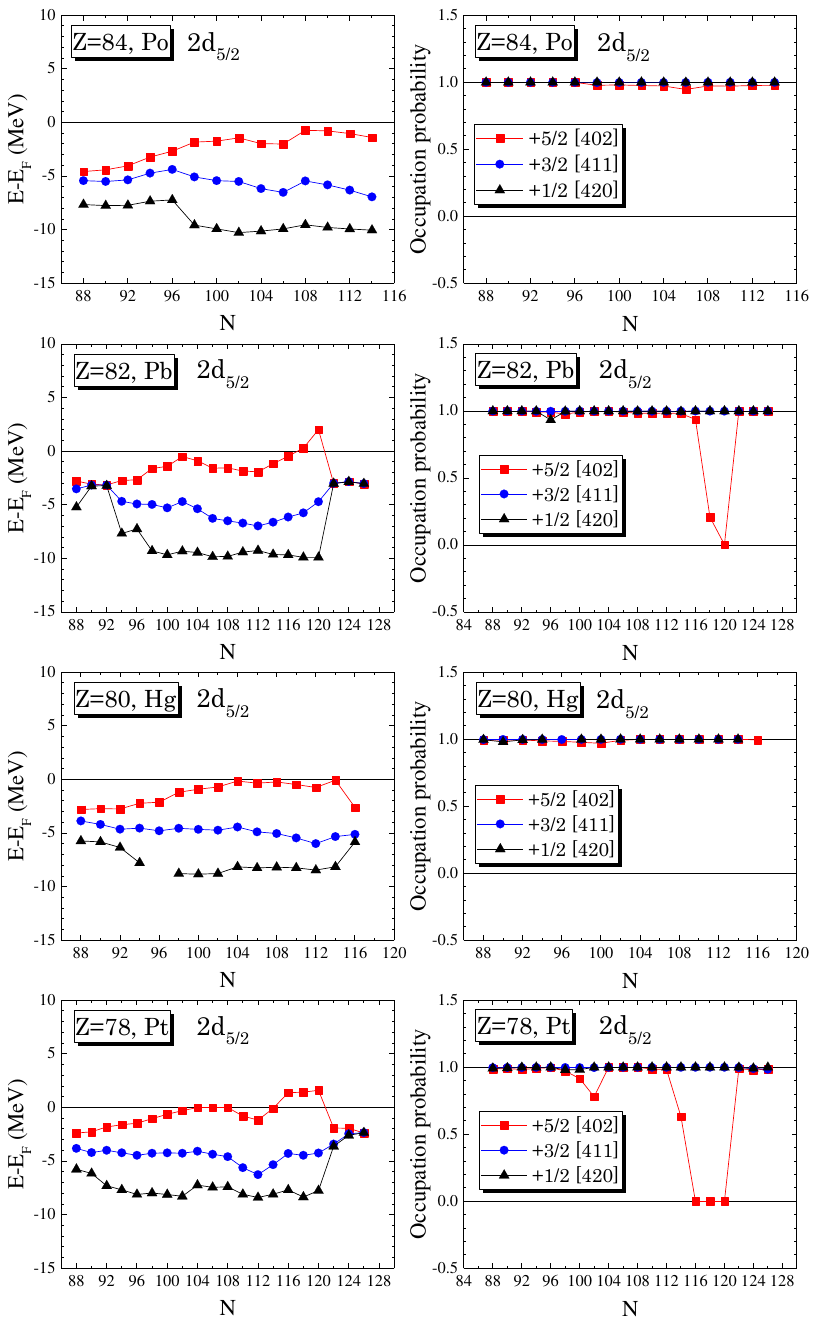}
     
    \caption{(Color online) Same as Fig. 1, but for the proton orbital $2d_{5/2}$. See Sec. III.A.1 for further discussion.}
        
\end{figure*}


\begin{figure*} [htb]

    \includegraphics[width=175mm]{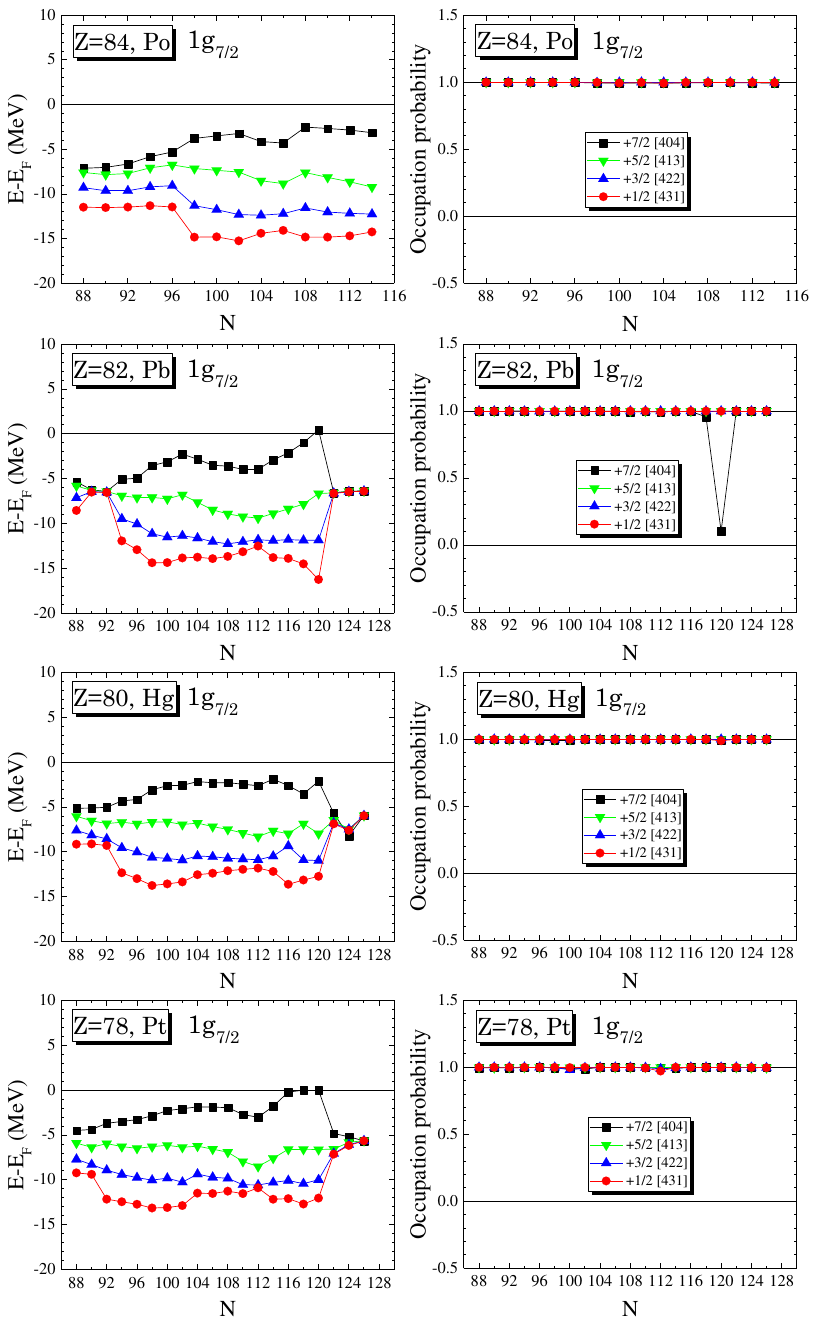}
     
    \caption{(Color online) Same as Fig. 1, but for the proton orbital $1g_{7/2}$. See Sec. III.A.1 for further discussion.}
     
\end{figure*}


\begin{figure*} [htb]

    \includegraphics[width=175mm]{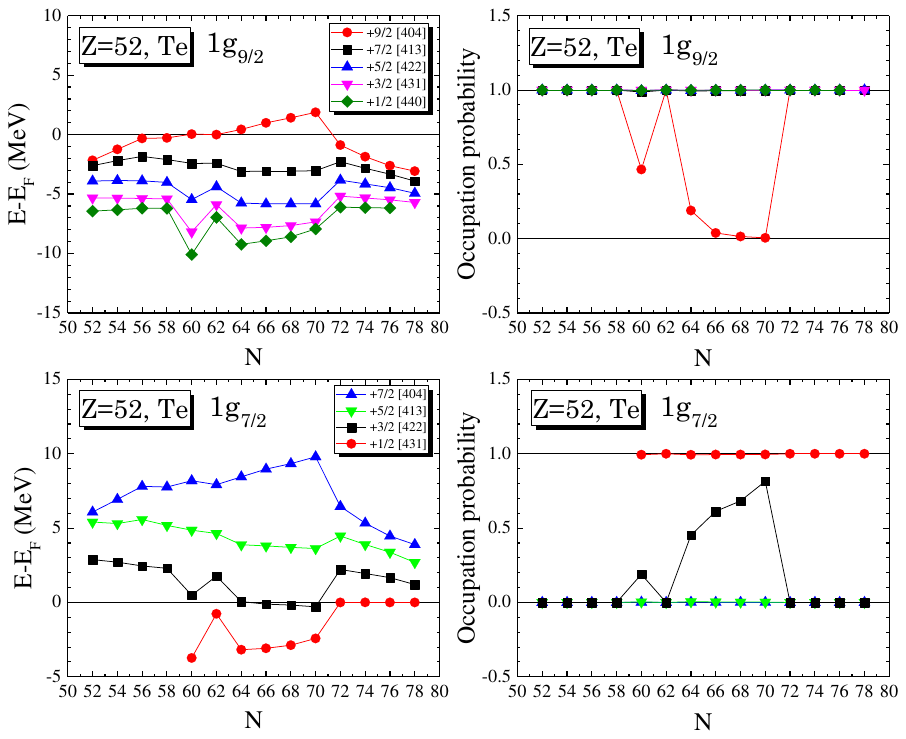}
     
    \caption{(Color online) Same as Fig. 1, but for the proton orbitals $1g_{9/2}$ and $1g_{7/2}$ for the Te isotopes. See Sec. III.A.2 for further discussion.}
  
\end{figure*}


\begin{figure*} [htb]

    \includegraphics[width=175mm]{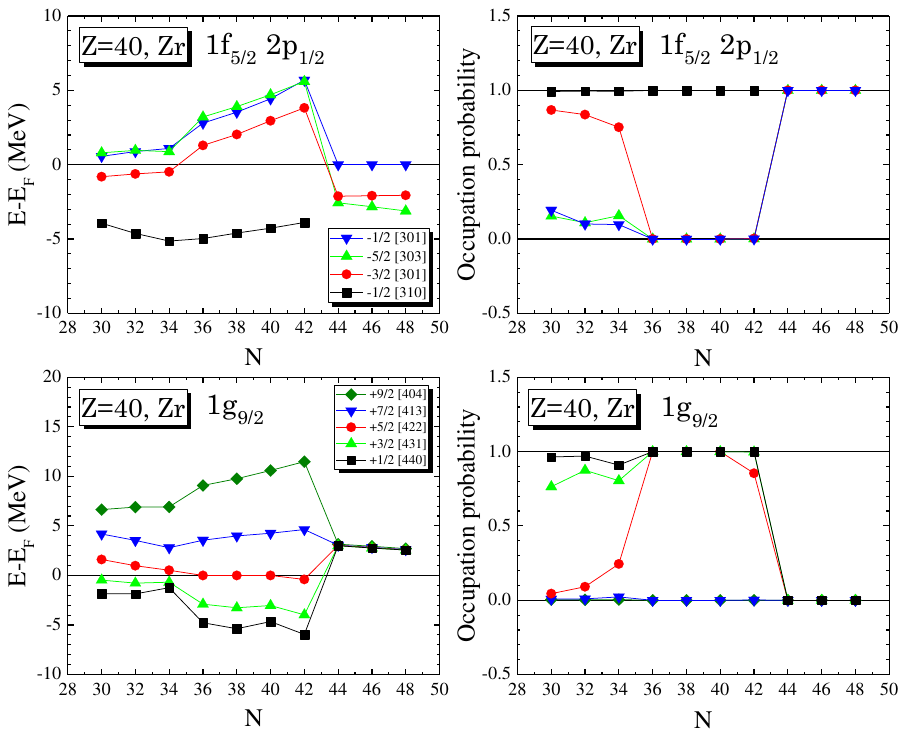}
     
    \caption{(Color online) Same as Fig. 1, but for the proton orbitals $1f_{5/2}$,  $2p_{1/2}$, and $1g_{9/2}$
     for the Zr isotopes. See Sec. III.A.3 for further discussion.}
    
\end{figure*}


\begin{figure*} [htb]

    \includegraphics[width=175mm]{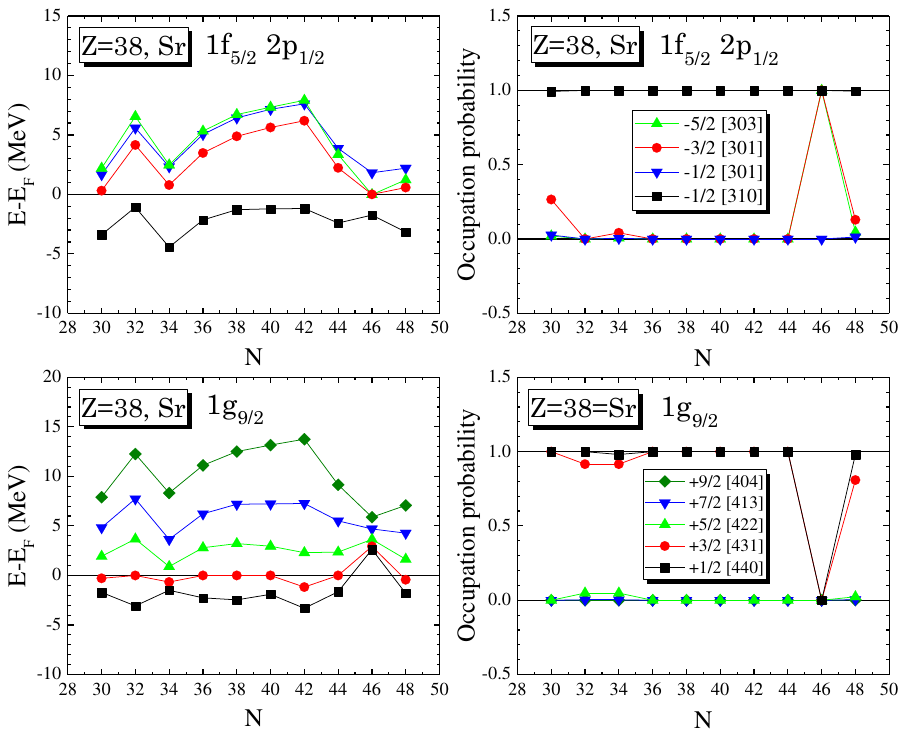}
     
    \caption{ (Color online) Same as Fig. 1, but for the proton orbitals $1f_{5/2}$,  $2p_{1/2}$, and $1g_{9/2}$
     for the Sr isotopes. See Sec. III.A.3 for further discussion.}
     
\end{figure*}


\begin{figure*} [htb]

    \includegraphics[width=175mm]{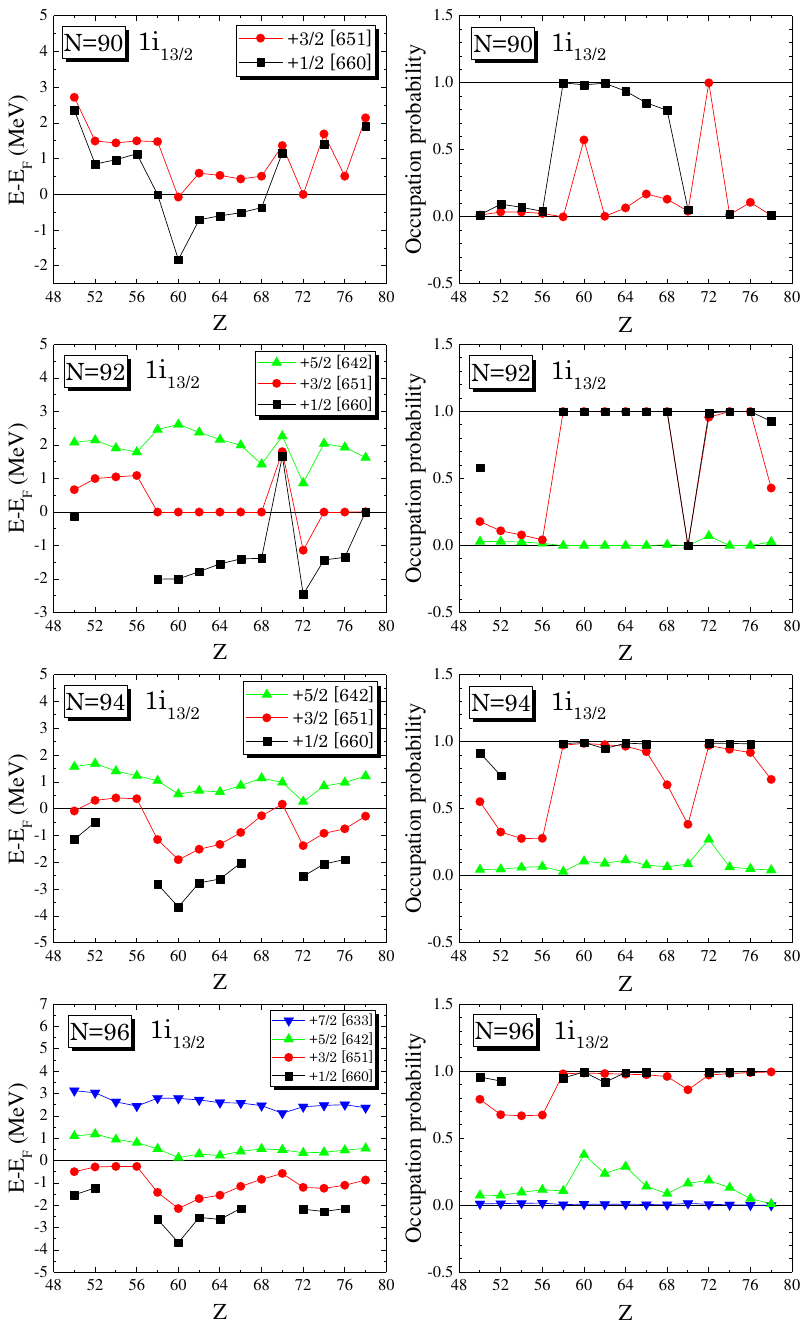}
     
    \caption{ (Color online) Left column: single-particle energy of neutron orbital $1i_{13/2}$ relative to the Fermi energy as a function of the proton number $Z$ obtained by non-relativistic mean-field HFB theory with Skyrme-SKI3 functional for $N=90$-96 isotones. Right column: occupation probability of neutron orbital $1i_{13/2}$ for $N=90$-96 isotones. See Sec. III.B.1 for further discussion.}
     
\end{figure*}


\begin{figure*} [htb]

    \includegraphics[width=175mm]{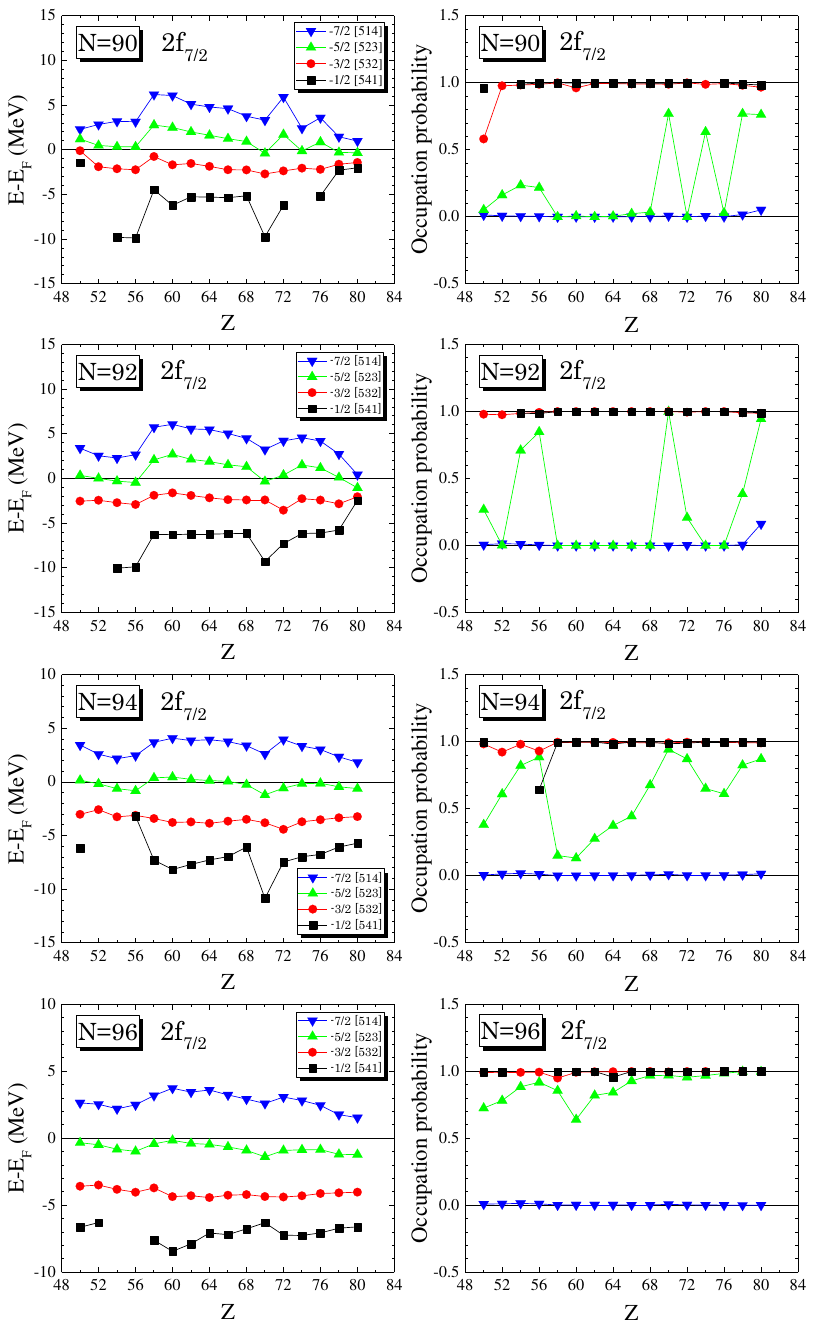}
    
        \caption{(Color online)  Same as Fig. 9, but for the neutron orbital $2f_{7/2}$. See Sec. III.B.1 for further discussion.}
     
\end{figure*}


\begin{figure*} [htb]

    \includegraphics[width=175mm]{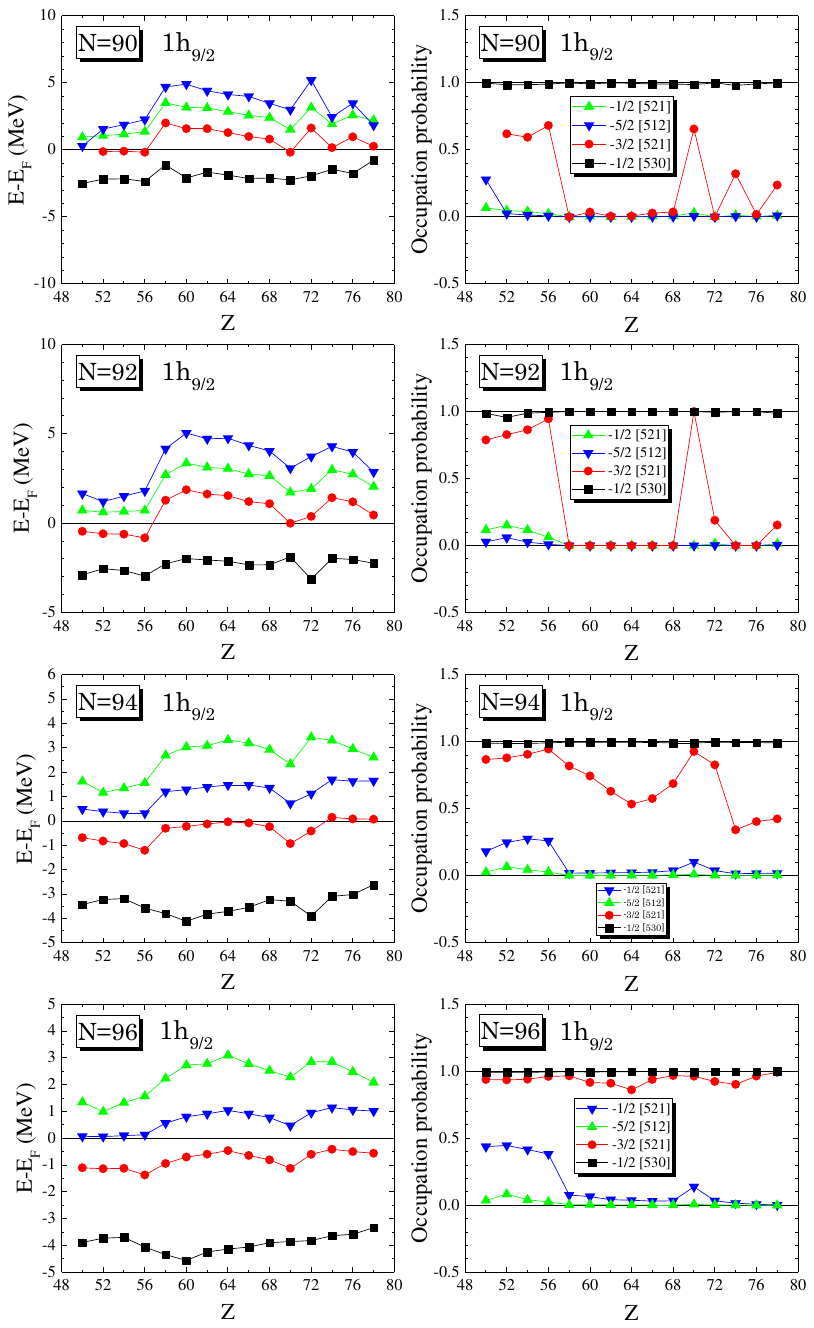}
     
    \caption{ (Color online) Same as Fig. 9, but for the neutron orbital $1h_{9/2}$. See Sec. III.B.1 for further discussion.}
     
\end{figure*}


\begin{figure*} [htb]

    \includegraphics[width=175mm]{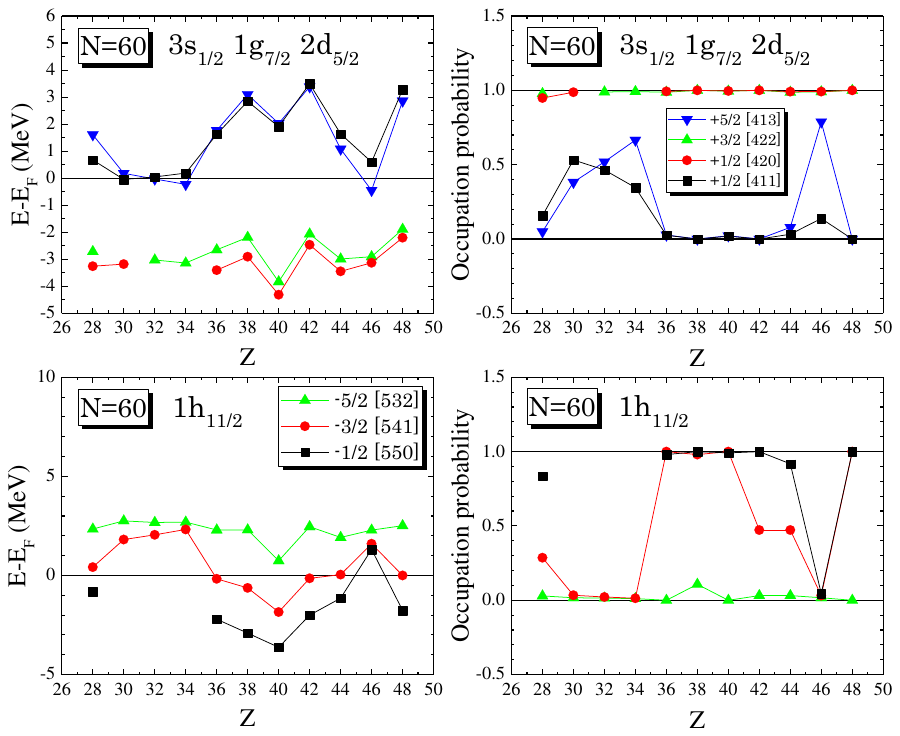}
     
    \caption{(Color online) Same as Fig. 9, but for the neutron orbitals $3s_{1/2}$, $1g_{7/2}$, $2d_{5/2}$, and $1h_{11/2}$ for the $N=60$ isotones. See Sec. III.B.2 for further discussion.}
     
\end{figure*}


\begin{figure*} [htb]

    \includegraphics[width=175mm]{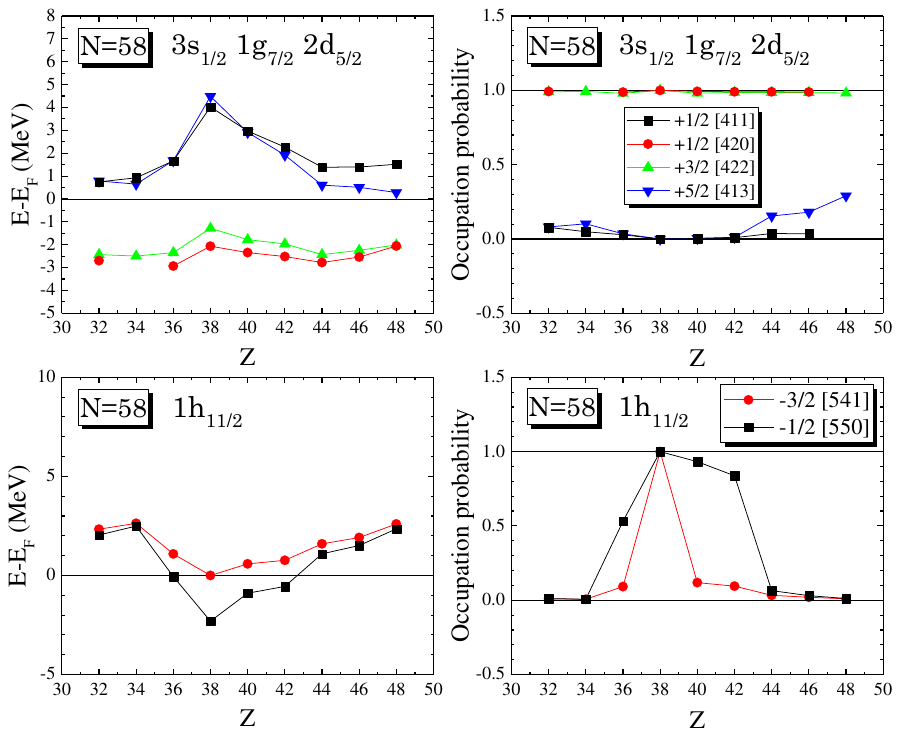}
     
    \caption{(Color online) Same as Fig. 9, but for the neutron orbitals $3s_{1/2}$, $1g_{7/2}$, $2d_{5/2}$, and $1h_{11/2}$ for the $N=58$ isotones. See Sec. III.B.3 for further discussion.}
    
\end{figure*}


\begin{figure*} [htb]

    \includegraphics[width=175mm]{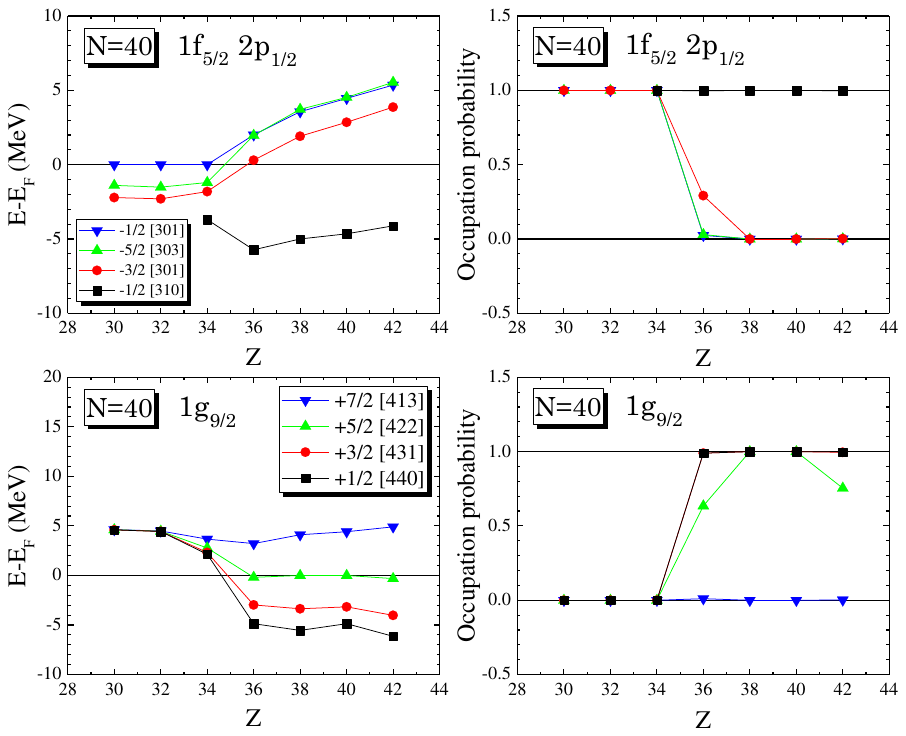}
     
    \caption{(Color online) Same as Fig. 9, but for the neutron orbitals $1f_{5/2}$, $2p_{1/2}$ and $1g_{9/2}$ for the $N=40$ isotones. See Sec. III.B.4 for further discussion.}
     
\end{figure*}


\begin{figure*} [htb]

    \includegraphics[width=175mm]{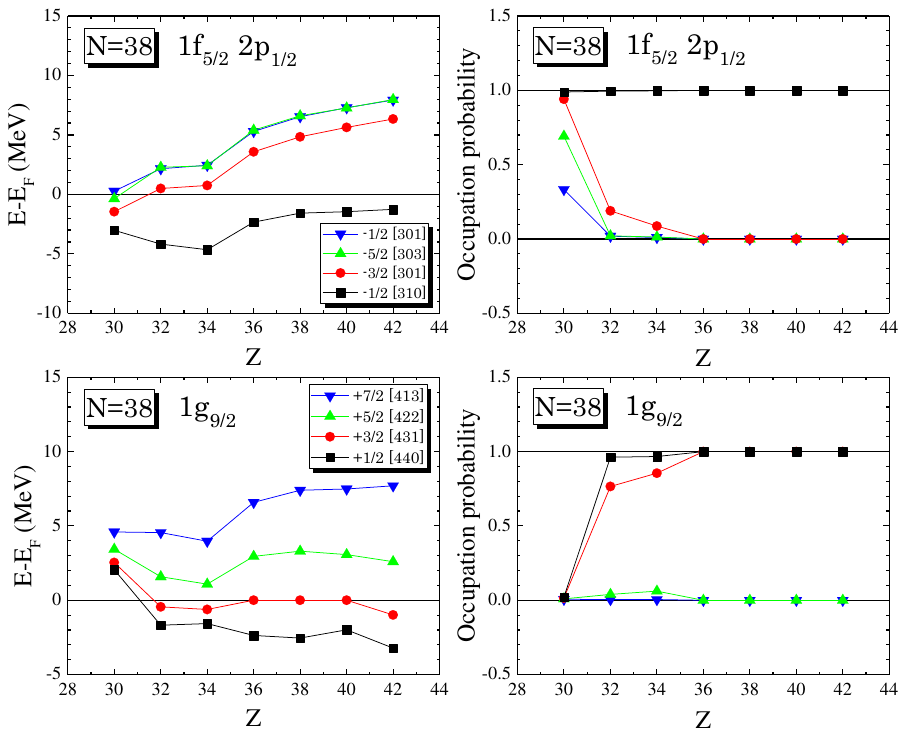}
     
    \caption{ (Color online) Same as Fig. 9, but for the neutron orbitals $1f_{5/2}$, $2p_{1/2}$ and $1g_{9/2}$ for the $N=38$ isotones. See Sec. III.B.4 for further discussion.}
     
\end{figure*}


\begin{figure*} [htb]

\includegraphics[width=160mm]{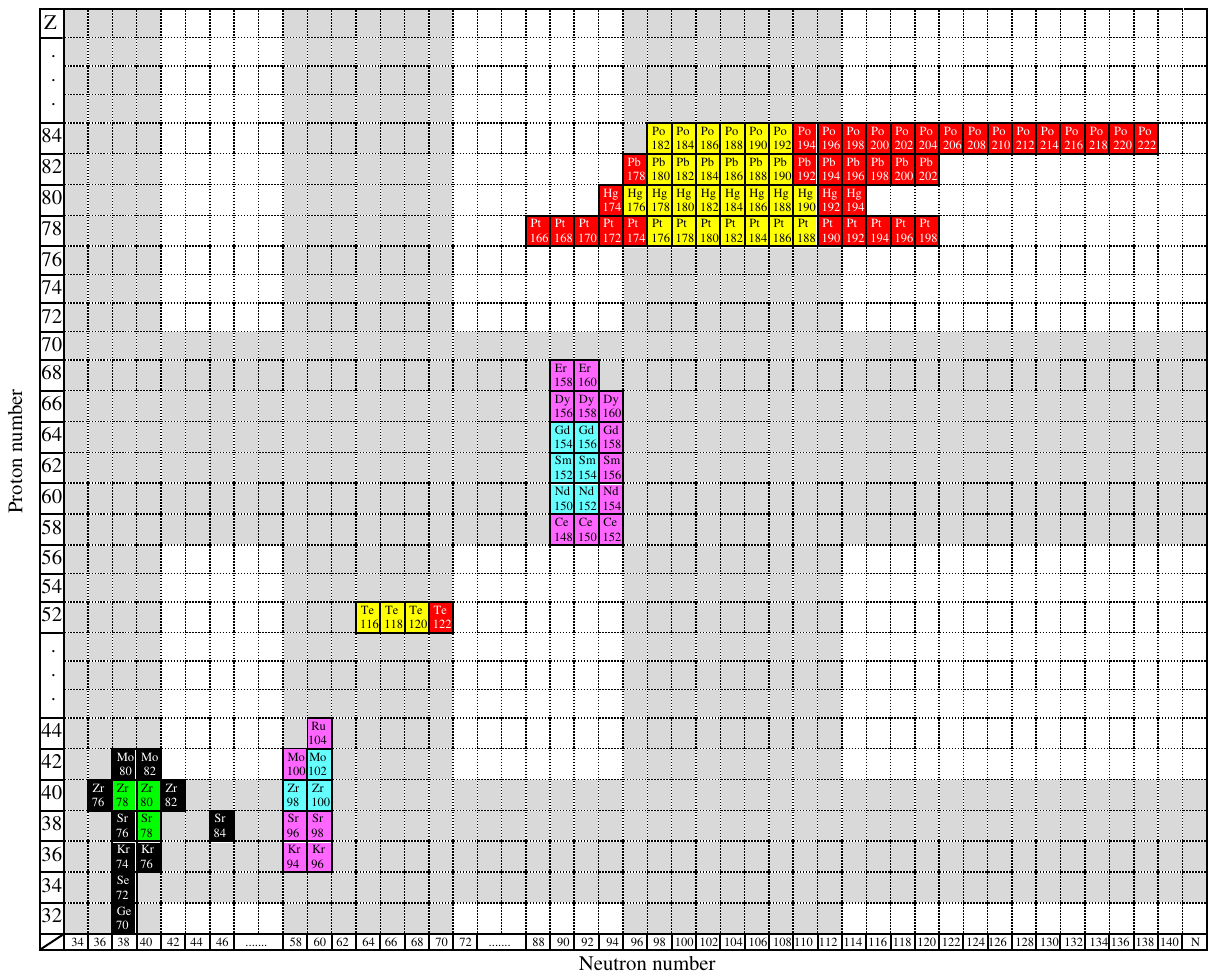}
     
\caption{ (Color online) Islands of shape coexistence (SC) found in the present study based on the particle-hole excitations mechanism in the single-particle energy of protons and neutrons relative to the Fermi energy obtained by non-relativistic mean-field HFB theory using Skyrme-SKI3 functional are compared with the islands of SC of Ref. \cite{Bonatsos2022b} using covariant density functional theory with DDME2 functional. Islands of SC predicted by both calculations are shown in yellow (neutron-induced islands), cyan (proton-induced islands), and green (islands induced by both mechanisms). In addition, islands of SC predicted only by the present Skyrme-SKI3 calculation are shown in red (neutron-induced islands), pink (proton-induced islands), and black (islands induced by both mechanisms). The gray stripes indicate the regions within which SC can be expected to occur according to the dual-shell mechanism within the Elliott SU(3) and the proxy-SU(3) symmetry \cite{Martinou2021}. As it is seen, most of the present predictions lie within the gray stripes, with the notable exception of many neutron-rich Pt, Pb, and Po isotopes, as well as with a few neutron-deficient Pt isotopes. See Section IV for further discussion. }
     
\end{figure*}

\end{document}